\documentclass[
  aps,
  prl,
  reprint,
  amsmath,
  amssymb,
  nofootinbib,
  superscriptaddress
]{revtex4-2}

\usepackage{bm}
\usepackage{mathtools}
\usepackage{amsthm}
\usepackage{microtype}
\usepackage{graphicx}
\usepackage{xcolor}
\usepackage{hyperref}

\hypersetup{
  colorlinks=true,
  linkcolor=blue!45!black,
  citecolor=blue!45!black,
  urlcolor=blue!45!black
}

\newcommand{\CC}{\mathbb C}

\newcommand{\dd}{\,\mathrm d}
\newcommand{\Gp}{G_{+}}
\newcommand{\Gm}{G_{-}}

\newcommand{\Log}{\mathrm{Log}}

\begin{document}

\title{Kinetic heterogeneity tightens thermodynamic bounds on spectra of Markov cycles}

\author{Rongxing Xu}
\email{xurongxing@uestc.edu.cn}
\affiliation{Institute of Fundamental and Frontier Sciences, University of Electronic Science and Technology of China, Chengdu, Sichuan 611731, China.}

\author{Cuiling Meng}
\email{cuilingmeng@uestc.edu.cn}
\affiliation{Institute of Fundamental and Frontier Sciences, University of Electronic Science and Technology of China, Chengdu, Sichuan 611731, China.}

\date{\today}

\begin{abstract}
Thermodynamic bounds on the spectrum of a Markov cycle are determined by quantities such as the cycle affinity and maximal escape rate, but do not distinguish cycles with different local rate profiles. In this work, we show that kinetic heterogeneity, through the geometric means of forward ($G_+$) and backward ($G_-$)rates, tightens the bound on oscillatory relaxation. Specifically, we prove that, for each winding sector of the eigenmodes, the oscillation frequency cannot exceed that of the uniform comparison cycle with rates $G_+$ and $G_-$. Combining with existing affinity-winding bounds, we also demonstrate a contracted region for the eigenvalues and shows that the boundary of Uhl-Seifert ellipse can be saturated if and only if cycles are uniform. Our proof provides a complex-analytic framework that unifies the kinetic and thermodynamic constraint.
\end{abstract}

\maketitle

\paragraph{Introduction---.}
Nonequilibrium systems such as chemical reaction networks \cite{schnakenberg1976network, gaspard2002correlation, PhysRevE.74.010902}, biomolecular machines \cite{qian2003thermodynamic, PhysRevLett.110.248102, mehta2012energetic, brown2017allocating}, and social processes \cite{liggett2013stochastic} are usually modeled as continuous-time Markov jump dynamics on a network of mesoscopic states. Their dynamical property are governed by the spectrum of the Markov generator. The zero eigenvalue corresponds to the stationary state, while the real parts of nonzero eigenvalues determine relaxation rates and complex-conjugate eigenvalue pairs generate damped oscillations. The spectrum and its localization have been shown to plays key roles in the oscillatory properties of relaxation dynamics \cite{PhysRevA.98.042118, 10.1063/1.532394, PhysRevLett.130.230404}, structure of correlation functions \cite{PhysRevLett.131.077101, PhysRevE.108.L042103, PhysRevResearch.6.013273} and degree of irreversibility \cite{PhysRevLett.114.158101, PhysRevE.106.014106, PhysRevE.105.064101, PhysRevResearch.6.013082}.

Recent studies show that the spectrum of the generator is constrained by both network structure and thermodynamics of the underlying stochastic dynamics \cite{Gershgorin1931, Dmitriev1946, mashreghi2007conjecture, kellogg1978complex, PhysRevE.95.062409, uhl2019affinity, PhysRevResearch.6.013082, z4t2-18cx, kolchinsky2026cycle}. For biochemical oscillators, the oscillatory frequency was shown to be bounded by thermodynamic affinity which measures the degree of nonequilibrium in a system that provides a driving force for equilibration \cite{PhysRevE.95.062409}. Uhl and Seifert then conjectured that the complete spectrum of a unicyclic generator lies inside an affinity-dependent ellipse whose boundary is traced by a uniform asymmetric random walk \cite{uhl2019affinity}. More general thermodynamic constraints on relaxation and oscillation were later derived from spectral perturbation and geometric properties of Markov rate matrices \cite{PhysRevResearch.6.013082, z4t2-18cx}. Most recently, cycle affinity together with the winding structure of eigenmodes was shown to localize complex eigenvalues more finely and to establish the Uhl-Seifert conjecture \cite{kolchinsky2026cycle}. These results provide universal bounds in terms of the cycle size, affinity, maximal escape rate, and mode winding.

However, these thermodynamic bounds are usually universal and do not distinguish different rate profiles that share these quantities. In other words, a uniform cycle and a strongly heterogeneous cycle can obey the same thermodynamic spectral bound even though their dynamics are substantially different. At the opposite extreme, the exact spectrum is obtained only when the full rate matrix is known. This leaves an unresolved intermediate level between localization only by thermodynamic quantities and complete microscopic details. How does limited information about the kinetic structure of a particular cycle further constrain its spectrum?

\begin{figure}[tbp]
    \centering
    \includegraphics[width=1\columnwidth]{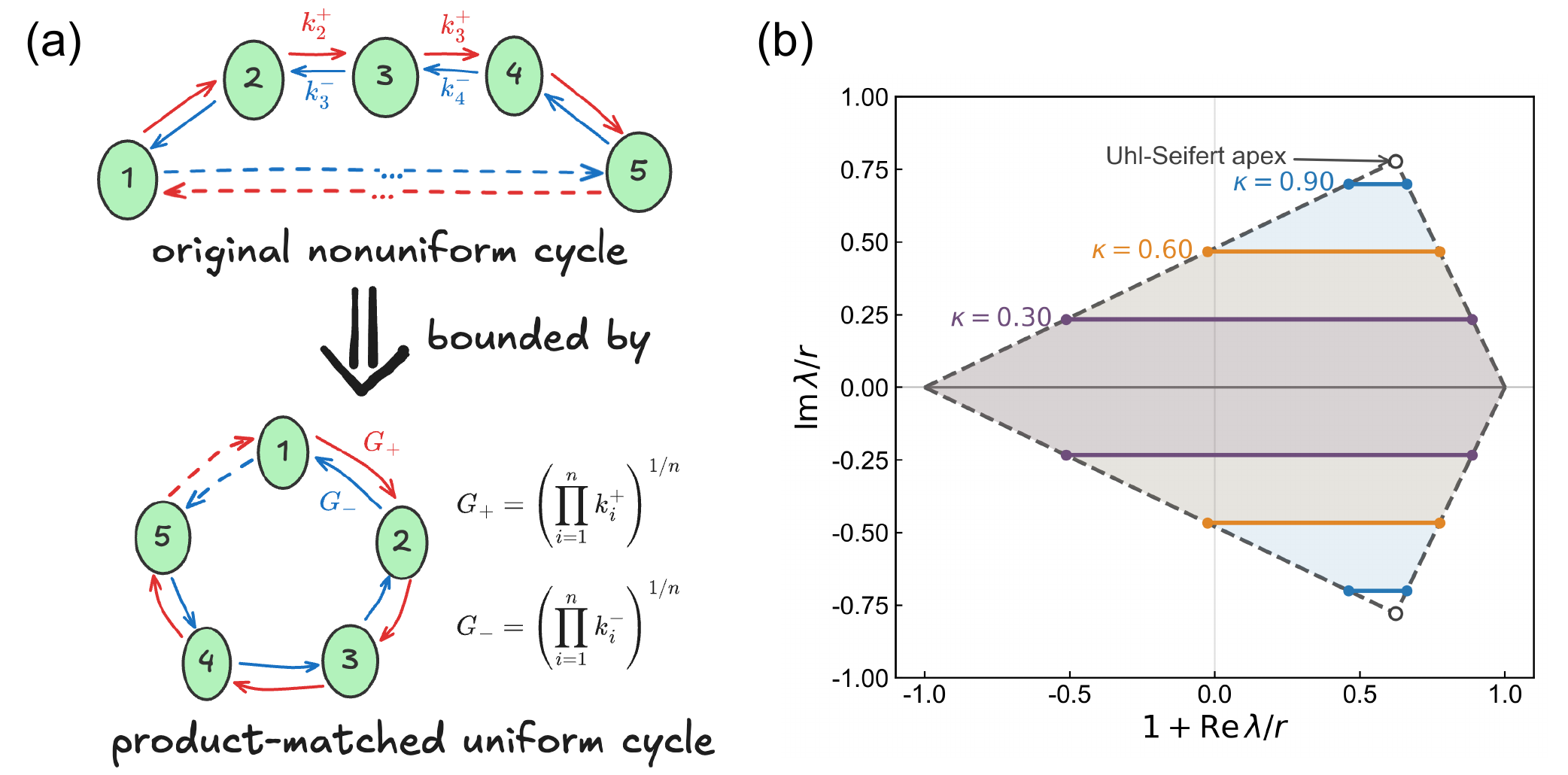}
    \caption{ (a) Schematic of original cycle and the comparison uniform cycle. An original nonuniform Markov cycle with local forward and backward rates $k_i^+$ and $k_i^-$ is associated with a product-matched uniform cycle whose rates are the geometric means $G_\pm$. For each winding sector, the oscillation frequency of the original cycle is bounded by that of this uniform comparison cycle. (b) Illustration of the additional localization supplied by the kinetic factor $\kappa$ at $n = 7$, $\mathcal A = 3$, $r = 1$, and $w = 1$. The dashed boundary is the affinity-winding bounds given by the Ref.\cite{kolchinsky2026cycle}, whose upper apex is on the Uhl-Seifert ellipse \cite{uhl2019affinity}. The horizontal kinetic caps are shown for $\kappa=0.90$, $0.60$, and $0.30$ respectively. Decreasing $\kappa$ leaves the affinity-winding bounds unchanged but removes its apex, yielding increasingly restrictive kinetic-refined regions.
}
\label{fig:1}
\end{figure}

In this work, we show that the geometric means of the forward and backward transition rates, $G_+$ and $G_-$, provide such a tightened constraint. We define a uniform comparison cycle with rates $G_+$ and $G_-$, which thus has the same cycle affinity as the original process (as shown in FIG.~\ref{fig:1}~(a)). We prove that, in every winding sector of eigenmodes, the oscillation frequency of the original cycle cannot exceed that of the corresponding mode of this comparison cycle. For every nonuniform cycle, the resulting frequency bound is strictly stronger than the recent established affinity-winding bounds \cite{kolchinsky2026cycle}. Combining with the existing affinity-winding bounds, it yields a refined ellipse for the complete spectrum (FIG.~\ref{fig:1}~(b)) and shows that the nonreal Uhl-Seifert boundary can be saturated only by uniform cycles. Our results establish a genuine intermediate level of spectral description and provide a unified routine to derive various existing bounds \cite{PhysRevE.106.014106, zheng2024topological, uhl2019affinity, PhysRevE.108.L042103, PhysRevResearch.6.013273}. Besides, our proof also provides an alternative analytic route to the Uhl-Seifert conjecture through a complex-analytic representation of the spectral problem. It places the kinetic and thermodynamic bounds within the same analytic structure.

\paragraph{Setup---.}

We start from a continuous-time Markov jump process on a ring of $n\geq3$ states (FIG.~\ref{fig:1}~(a)). From state $i$, the process jumps to $(i+1)$ with rate $k_i^+>0$ and to $(i-1)$ with rate $k_i^->0$. Let $\bm p(t)$ denote the probability distribution. Its evolution obeys the master equation: ${\dd\bm p}/{\dd t}=\bm R^{\mathsf T}\bm p$, where $\bm R$ is the irreducible generator (also known as the rate matrix). With $R_{ji}$ denoting the transition rate from state $j$ to state $i$, its elements are written by
\begin{equation}
    R_{ji} = k_j^+ \delta_{i,j+1} + k_j^- \delta_{i,j-1} - (k_j^+ + k_j^-) \delta_{ij}.
\end{equation}
The diagonal terms define the local escape rates $r_i=-R_{ii}=k_i^++k_i^-$, and we write the maximum escape rates by $r=\max_i r_i$.

In this work, we focus on the oscillatory relaxation, which requires broken detailed balance. The thermodynamic bias accumulated during one forward traversal is the cycle affinity, defined by
\begin{equation}
    \mathcal A=\ln\frac{\prod_{i=1}^{n}k_i^+}{\prod_{i=1}^{n}k_i^-}.
\end{equation}
Under local detailed balance, $\mathcal A$ is the entropy transferred to the reservoirs, in units of $k_{\mathrm B}$, by one net forward cycle. We orient the ring so that $\mathcal A>0$ and set $\alpha=\mathcal A/(2n)$. When $\mathcal A=0$, the unicyclic system satisfies detailed balance condition and is diagonally similar to a real symmetric matrix, so its spectrum is real \cite{schnakenberg1976network, seifert2012stochastic}. Thus, nonzero affinity is necessary for oscillatory eigenmodes.

The affinity fixes only the ratio of the forward and backward rate products. Consider the forward and backward cycle geometric means of transition rates:
\begin{equation}
    \Gp=\left(\prod_{i=1}^{n}k_i^+\right)^{1/n},
    \qquad
    \Gm=\left(\prod_{i=1}^{n}k_i^-\right)^{1/n}.
\end{equation}
By construction, $\Gp/\Gm=\exp(\mathcal A/n)=\exp(2\alpha)$. Once the affinity is fixed, $\Gp$ and $\Gm$ contain only one additional independent scalar, which may be taken as their sum $\Gp+\Gm$. The pair $(\Gp,\Gm)$ also defines a natural uniform comparison cycle, with uniform forward and backward rates $k_i^{+,\mathrm g}=\Gp$ and $k_i^{-,\mathrm g}=\Gm$ respectively. This comparison cycle has the same forward and backward rate products, and therefore the same cycle affinity, as the original process (FIG.~\ref{fig:1}~(a)).

We next characterize the phase structure of the oscillatory modes. Let $\lambda$ be an eigenvalue of $\bm R$ with $\operatorname{Im}\lambda>0$, and let $\bm f\in\CC^n$ be a corresponding right eigenvector, satisfying ${\bm R}{\bm f}=\lambda{\bm f}$. Its winding number is then defined by
\begin{equation}
    w(\bm f)=\frac{1}{2\pi}\sum_{i=1}^{n}\operatorname{Arg}\!\left(\frac{f_{i+1}}{f_i}\right),
    \qquad f_{n+1}=f_1,
    \label{eq:winding}
\end{equation}
where $\operatorname{Arg}z$ denotes the phase of $z$ chosen in the range $(-\pi,\pi]$. As shown in the Supplemental Material, the components of $\bm f$ do not vanish, the adjacent phase increments can be chosen in $(0,\pi)$, and $w$ is an integer satisfying $1\leq w<n/2$. We denote $\theta_w=2\pi w/n$. Physically, $w$ measures how many times the phase of an oscillatory eigenmode winds by $2\pi$ as one moves once around the state cycle. For a uniform cycle, it reduces exactly to the Fourier mode index. 

For later comparison, we recall the affinity-winding frequency bound derived by the consequence of Ref.~\cite{kolchinsky2026cycle}. Every eigenvalue with winding number $w$ satisfies
\begin{equation}
    \operatorname{Im}\lambda \leq r\tanh\alpha\,\sin\theta_w.
    \label{eq:universal_frequency_bound}
\end{equation}
The right-hand side depends only on the cycle size $n$, the affinity $\mathcal A$, the maximal escape rate $r$, and the winding number $w$. It is identical for all generators sharing these quantities, irrespective of their remaining kinetic structure. 

\paragraph{Main results---.}

We now state our first main result. Its sketched proof is given in the next section, with the remaining analytic details deferred to the Supplemental Material.

\paragraph{Theorem 1 (Geometric-mean frequency bound).}
Let $\lambda$ be an eigenvalue of $\bm R$ with $\operatorname{Im}\lambda > 0$ and winding number $w$. Then
\begin{equation}
    \operatorname{Im}\lambda \leq (\Gp-\Gm)\sin\theta_w.
    \label{eq:geometric_mean_bound}
\end{equation}
The bound is tight. Equality is attained by the product-matched uniform cycle with forward and backward rates $k_i^{+,\mathrm g}=\Gp$ and $k_i^{-,\mathrm g}=\Gm$.

The interpretation of Eq.~\eqref{eq:geometric_mean_bound} is immediate. The eigenvalue with winding number $w$ of the uniform comparison cycle is \cite{uniform_cycle_eigenvalue}:
\begin{equation}
    \lambda_w^{\mathrm g} = -(\Gp+\Gm)(1-\cos\theta_w) + i(\Gp-\Gm)\sin\theta_w.
    \label{eq:comparison_eigenvalue}
\end{equation}
Hence, Theorem~1 can be written equivalently as $\operatorname{Im}\lambda \leq \operatorname{Im}\lambda_w^{\mathrm g}$. Among cycles with the same forward and backward rate products, the product-matched uniform cycle sets the largest winding-resolved oscillation frequency allowed by the bound. In other words, kinetic heterogeneity can slow down the oscillations. Even at the same total affinity, fast and slow jumps coexist, and the slow jumps hinder the uniform propagation of phase around the cycle.

\begin{figure}[tbp]
    \centering
    \includegraphics[width=0.6\columnwidth]{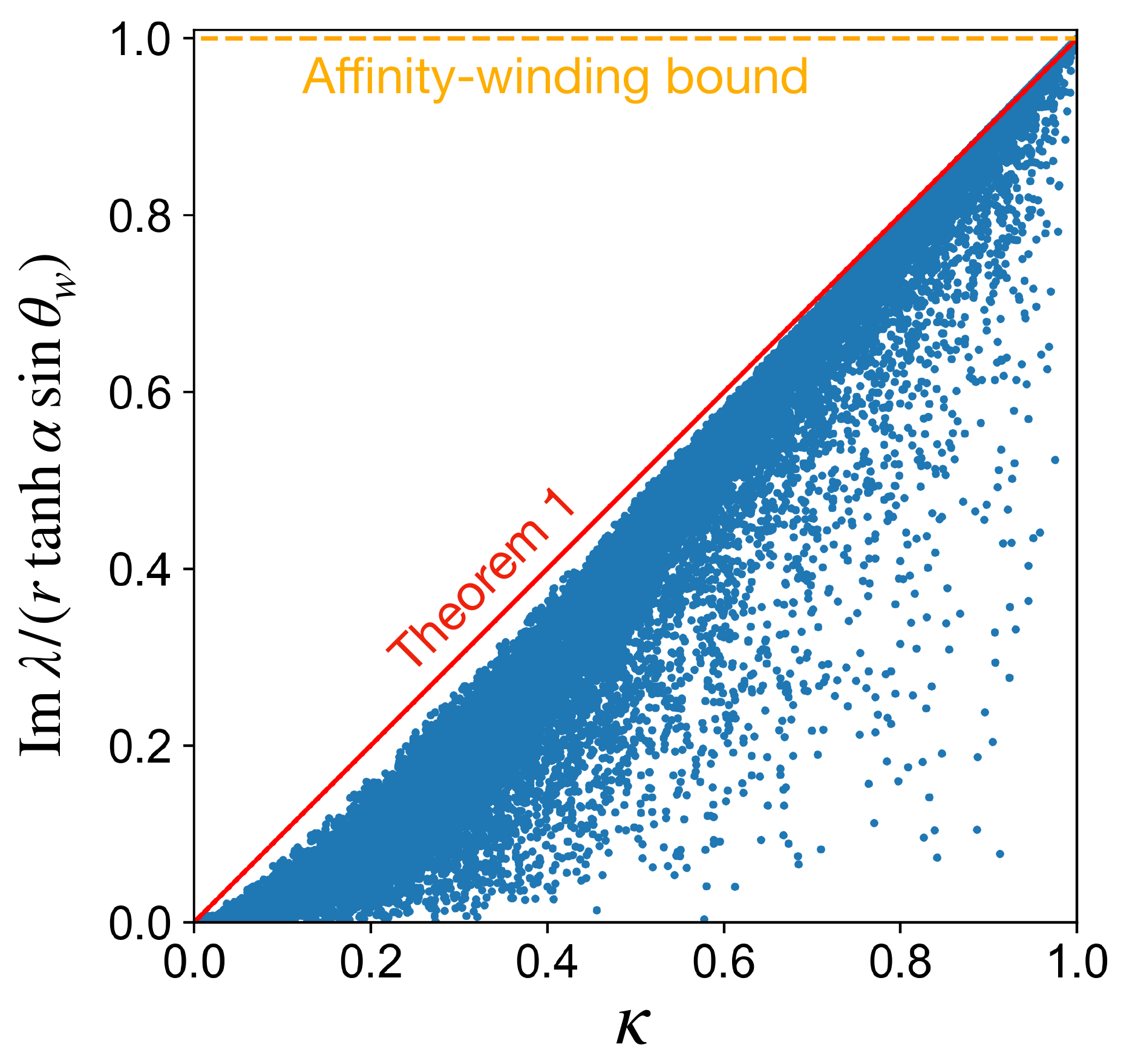}
    \caption{
    Numerical verification of the geometric-mean frequency bound for the cycle with $n=7$. Each point represents a nonreal eigenmode of a numerically generated nonuniform Markov cycle, plotted in terms of $\kappa$ and the normalized oscillation frequency. The red solid line is the geometric-mean bound given by Theorem~1, whereas the orange dashed line is the affinity-winding bound on the imaginary part of $\lambda$. 
}
\label{fig:2}
\end{figure}

To compare Eq.~\eqref{eq:geometric_mean_bound} directly with the bound \eqref{eq:universal_frequency_bound}, we introduce the dimensionless ratio
\begin{equation}
    \kappa = \frac{\Gp+\Gm}{r}.
    \label{eq:kappa}
\end{equation}
to quantify the heterogeneity of the rates. Since $\Gp/\Gm=\exp(2\alpha)$, we have $\Gp-\Gm=(\Gp+\Gm)\tanh\alpha=r\kappa\tanh\alpha$. One can prove that $0< \kappa \leq 1$ \cite{Holder}. Then, Theorem 1 gives
\begin{equation}
    \operatorname{Im}\lambda\leq r\kappa\tanh\alpha\sin\theta_w\leq r\tanh\alpha\sin\theta_w.
    \label{eq:frequency-hierarchy}
\end{equation}
The rightmost expression is precisely the bound \eqref{eq:universal_frequency_bound}. For every nonuniform cycle, $\kappa<1$, so the first inequality provides a strictly stronger frequency bound. The equality is achieved only for the uniform cycle. Figure~\ref{fig:2} numerically verifies Theorem~1 for randomly generated nonuniform cycles. All sampled eigenmodes satisfy our geometric-mean frequency bound, with the reduction from the affinity-winding bound controlled by $\kappa$.

In fact, $G_{\pm}$ do not specify the ordering or full distribution of the local transition rates. In principle, they can be measured from the short-time probability of a purely forward or backward cycle traversal, $P_\pm^{\rm cyc}(t)=(G_\pm t)^n/n!+O(t^{n+1})$. Theorem~1 is thus the consequence of coarse-grained kinetic heterogeneity, which sets an intermediate level between thermodynamics and complete microscopic details.

With Theorem 1, we can immediate obtain a contracted admissible region of the spectrum. For an eigenvalue $\lambda$ with winding number $w$, define the normalized spectral coordinates $(X,Y)=(1+{\operatorname{Re}\lambda}/{r}, {|\operatorname{Im}\lambda|}/{r})$, and introduce $(U_w, V_w) = (\cos\theta_w, \tanh\alpha\,\sin\theta_w)$. In Ref.~~\cite{kolchinsky2026cycle}, they established the slow and fast-decay affinity-winding bounds:
\begin{equation}
    Y \leq V_w\frac{1-X}{1-U_w} ~~{\rm and}~~
    Y \leq V_w\frac{1+X}{1+U_w}.
    \label{eq:affinity_winding_bounds}
\end{equation}
The first implies that fast oscillation, slow decay, and high frequency require large affinity. The second indicates that fast oscillation, fast decay, and low frequency also require large affinity. For fixed $w$, these two inequalities define a triangle with vertices $(-1,0)$, $(1,0)$, and $(U_w,V_w)$ (Fig.~\ref{fig:1} (b)). The apex $(U_w,V_w)$ is precisely the winding-$w$ eigenvalue of the uniform cycle after normalization which should on the Uhl-Seifert conjecture.

Here, Theorem~1 adds an additional horizontal constraint $Y \leq \kappa V_w$ on the two affinity-winding bounds as shown in FIG.~\ref{fig:1} (b). The joint consequence of Eqs.~\eqref{eq:affinity_winding_bounds} and Theorem 1 is the following.

\paragraph{Corollary 1 (Kinetic refinement of winding-resolved spectral region).}
Every eigenvalue with winding number $w$ satisfies
\begin{equation}
    Y \leq \min
    \left\{
        V_w\frac{1-X}{1-U_w},
        \;
        V_w\frac{1+X}{1+U_w},
        \;
        \kappa V_w
    \right\}.
    \label{eq:trapezoid_bound}
\end{equation}

For $\kappa=1$, the horizontal bound passes through the apex $(U_w,V_w)$ and Eq.~\eqref{eq:trapezoid_bound} reduces to the affinity-winding triangle. For every nonuniform cycle, $\kappa<1$, and the apex is cut off. The admissible region is then the trapezoid with upper vertices $L_{w,\kappa} = \left( \kappa-1 + \kappa U_w, \kappa V_w \right)$ and $ R_{w,\kappa} = \left( 1-\kappa + \kappa U_w, \kappa V_w \right)$. Thus, the affinity controls the trade-off between decay and oscillation through the two oblique boundaries, while kinetic heterogeneity imposes an kinetic cap on the oscillation frequency.

Equation~\eqref{eq:trapezoid_bound} is the strongest joint constraint here, since it retains $w$ explicitly. To compare with the Uhl-Seifert ellipse, we try to discard $w$ and seek a single centered ellipse 
\begin{equation}
    X^2+\frac{Y^2}{\tanh^2\alpha\,C}\leq1
    \label{eq:generic_ellipse}
\end{equation}
containing all these winding-resolved trapezoids. According to Eq.~\eqref{eq:trapezoid_bound}, the contraction factor $C$ reads $C_w(\kappa) = {\kappa(1+|U_w|)}/{\left[2-\kappa(1-|U_w|)\right]}$. Since $C_w$ increases monotonically with $|U_w|$, the weakest common constraint over all $w$ is obtained at
\begin{equation}
    \rho_n = \max_{1\leq w<n/2}|U_w|
    = \begin{cases}
        \cos(\pi/n), & n\ {\rm odd},\\[2mm]
        \cos(2\pi/n), & n\ {\rm even}.
    \end{cases}
    \label{eq:rho_n}
\end{equation}
where the parity dependence follows from the discrete set of nonreal winding-$w$ modes. The resulting winding-independent contraction factor is $C_n(\kappa) = {C_w(\kappa; \rho_n)}$.

\begin{figure}[tbp]
    \centering
    \includegraphics[width=1\columnwidth]{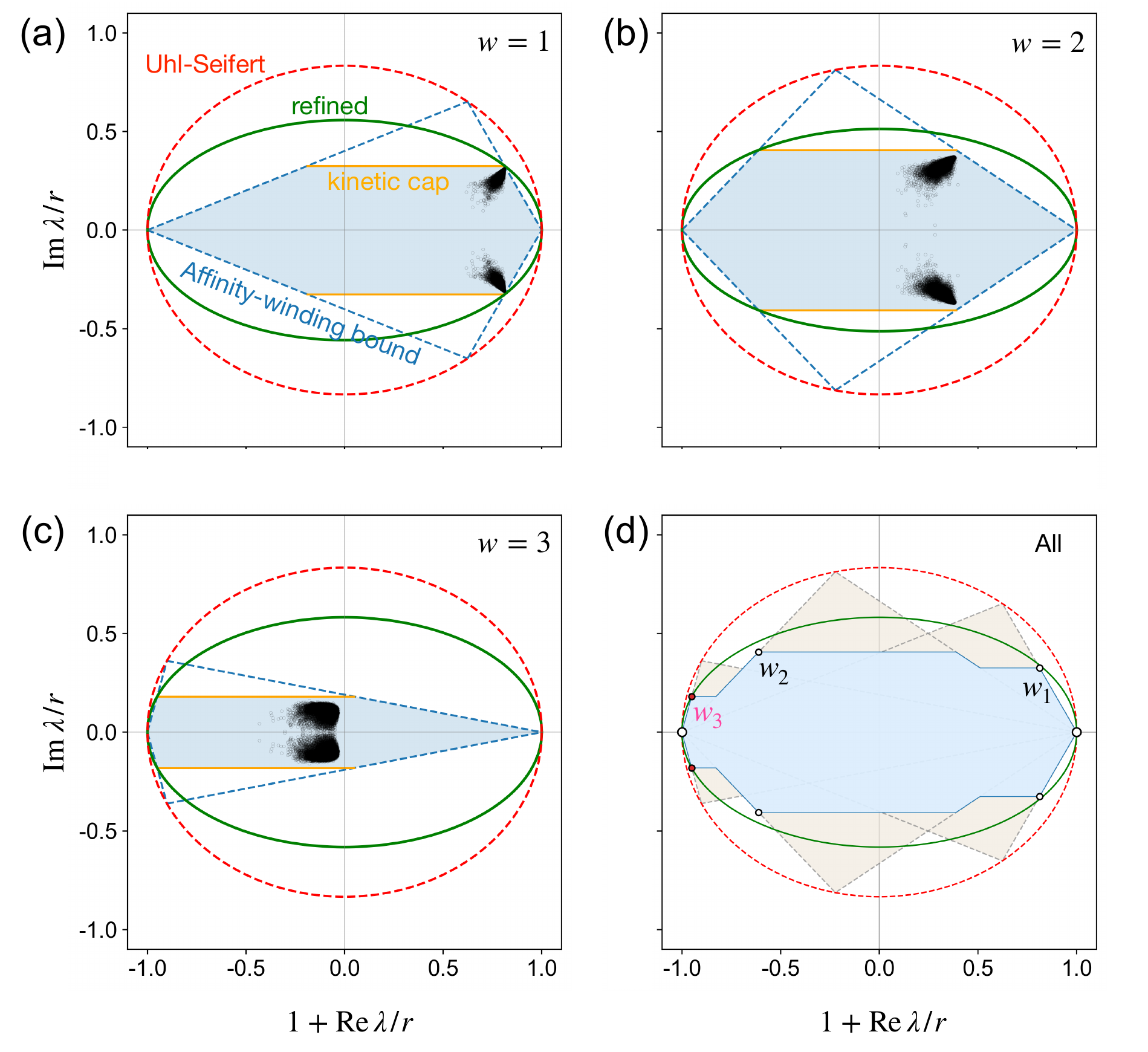}
    \caption{
    Kinetic refinement of winding-resolved and winding-independent spectral localization for fixing $n=7$, $\kappa = 0.5$ and ${\alpha} = 1.2$. (a)--(c) Spectral localization for winding sectors $w=1$, $2$, and $3$, respectively. The blue dashed lines denote the established affinity-winding bounds, while the orange horizontal lines show the additional kinetic cap from Theorem~1. Their intersection gives the shaded winding-resolved region, obtained by truncating the apex of the universal affinity-winding triangle. The black points are numerically computed eigenvalues in the corresponding winding sector. The red dashed curve is the Uhl-Seifert ellipse, whereas the green solid curve is the kinetic-refined elliptical boundary of Corollary~2. (d) Superposition of the winding-resolved refined regions. The labeled vertices indicate the kinetic caps associated with the different winding sectors. The winding-independent contraction factor $C_n(\kappa)$ is obtained at $w = 3$ (noted by red dots).
}
\label{fig:3}
\end{figure}

\paragraph{Corollary 2 (Kinetic-refined elliptical region).}
The complete spectrum of the generator is contained in the refined ellipse:
\begin{equation}
    \frac{(\operatorname{Re}\lambda+r)^2}{r^2} +
    \frac{(\operatorname{Im}\lambda)^2}
    {r^2\tanh^2\alpha\,C_n(\kappa)}
    \leq 1.
    \label{eq:refined_ellipse}
\end{equation}
For $\kappa=1$, one has $C_n(1)=1$, and Eq.~\eqref{eq:refined_ellipse} reduces to the Uhl-Seifert ellipse. For every nonuniform cycle, $\kappa<1$ and hence $C_n(\kappa)<1$, so this refined ellipse lies strictly inside the Uhl-Seifert ellipse away from the real axis.

Fig.~\ref{fig:3} provides a direct numerical verification of the above spectral bounds. For fixed $n$, the Uhl-Seifert ellipse has a vertical semiaxis $r\tanh\alpha$, determined by the affinity and maximal escape rate. The additional kinetic information enters through the contraction factor $\kappa$ (Fig.~\ref{fig:3} (a)--(c)). Eigenvalues for different winding sectors all remain inside the corresponding kinetic-refined polygons (lightblue region of Fig.~\ref{fig:3} (d)), and the complete spectrum is contained within the refined ellipse. The contraction also fixes the equality condition of the universal bound.

\paragraph{Corollary 3 (Rigidity of Uhl-Seifert ellipse saturation).}
A nonreal eigenvalue can lie on the boundary of the Uhl-Seifert ellipse, if and only if the cycle is uniform. Conversely, every nonreal eigenvalue of a uniform cycle lies on that boundary. Equivalently, nonreal boundary saturation occurs if and only if $k_i^+ = k^+$ and $k_i^- = k^-$. The stationary eigenvalue $\lambda = 0$, which is present for every Markov generator and lies at a real-axis endpoint of the ellipse, is excluded from this statement.

\paragraph{Sketch of the proof---.}

We now outline the sketch of the proof of Theorem~1. The proof of Corollaries are straight forward. We first convert the nonreciprocal eigenvalue problem of the Markov cycle into a reciprocal Floquet problem \cite{teschl2000jacobi}. Then, we encode the phase growth in an analytic self-map of the upper half-plane $\mathbb H$. Last, we show that Theorem 1 and affinity-winding bounds are obtained by considering the limit behavior of the map at $\pm 1$ and infinity. The detailed proof and the associated analytical derivations are given in the Supplemental Material.

We first normalize the generator by introducing $z=1+{\lambda}/{r}$, and regard the corresponding spectral parameter as an independent variable $\zeta \in \mathbb H$. A positive gauge transformation, $f_i=d_i v_i$ with ${d_{i+1}}/{d_i} = \sqrt{{k_{i+1}^-}/{k_i^+}}$, converts the eigenvalue equation of the generator into the reciprocal Jacobi recurrence
\begin{equation}
    c_i v_{i+1}(\zeta) + c_{i-1}v_{i-1}(\zeta) + s_i v_i(\zeta) = \zeta v_i(\zeta),
    \label{eq:jacobi_recurrence}
\end{equation}
with $c_i = {\sqrt{k_i^+k_{i+1}^-}}/{r}$ and $s_i = 1-{(k_i^++k_i^-)}/{r}$. The local forward-backward asymmetry has thus been removed from the off-diagonal couplings. It reappears globally because the gauge is not periodic: ${d_{i+n}}/{d_i} = e^{-\mathcal A/2}$. Since the original eigenvector is periodic, the gauged solution satisfies $v_{i+n}(z) = e^{\mathcal A/2}v_i(z) = e^{n\alpha}v_i(z)$. Thus the local nonreciprocity is replaced by a Floquet multiplier determined solely by the cycle affinity.

The full-cycle multiplier $e^{\mathcal A/2}$ does not retain the integer winding number, because the phase factor $e^{2\pi i w}$ disappears after one complete traversal. To recover the phase advance per transition, we select the expanding Floquet branch of Eq.~\eqref{eq:jacobi_recurrence}. As shown in the Supplemental Material, its adjacent ratios lie in the upper half-plane, i.e., ${v_{i+1}(\zeta)}/{v_i(\zeta)} \in\mathbb H$ and $\zeta\in\mathbb H$. It allows us to define the canonical per-step Floquet factor
\begin{equation}
    q(\zeta) =
    \exp\left[
        \frac{1}{n} \sum_{i=1}^{n}
        \Log_{\mathbb H}
        \left(
            \frac{v_{i+1}(\zeta)}{v_i(\zeta)}
        \right)
    \right],
    \label{eq:q_def}
\end{equation}
where $\Log_{\mathbb H}$ denotes the logarithm with argument in $(0,\pi)$ \cite{Logarithm}. The function $q$ is analytic in $\mathbb H$, satisfies $q(\mathbb H)\subset\mathbb H$ and $|q|>1$, and at the eigenvalue $z$ obeys $q(z)=e^{\alpha+i\theta_w}$. Thus $q(z)$ packages the affinity and the winding number into the modulus and phase of a single complex quantity. We can then naturally define the map $\Phi:\mathbb H\longrightarrow\mathbb H$:
\begin{equation}
    \Phi(\zeta) = \frac{q(\zeta)+q(\zeta)^{-1}}{2\cosh\alpha}.
    \label{eq:Phi_def}
\end{equation}
At the eigenvalue $z$, we have $\Phi(z) = \cos\theta_w + i\tanh\alpha\,\sin\theta_w$. Since $\Phi$ is an analytic self-map of $\mathbb H$, it admits the Herglotz-Nevanlinna representation \cite{adamyan2009modern}
\begin{equation}
    \Phi(\zeta) = \frac{\zeta}{\kappa} + \beta +
    \int_{\mathbb R}
    \left(
        \frac{1}{t-\zeta} - \frac{t}{1+t^2}
    \right)
    \dd\mu(t),
    ~{\rm where}~ \mu\geq0,
    \label{eq:nevanlinna_representation}
\end{equation}
with $\beta\in\mathbb R$. We now extract the spectral bounds from its limiting behavior at infinity and at the two real endpoints $\zeta=\pm1$.

We first consider $|\zeta| \to \infty$. Dividing Eq.~\eqref{eq:jacobi_recurrence} by $v_i$ gives
${v_{i+1}(\zeta)}/{v_i(\zeta)}={\zeta}/{c_i}+O(1)$, and hence
$q(\zeta)={\zeta}/{c_{\rm g}}+O(1)$ with $c_{\rm g}=\left(\prod_{i=1}^{n}c_i\right)^{1/n}={\sqrt{\Gp\Gm}}/{r}$. Because $2c_{\rm g}\cosh\alpha=(\Gp+\Gm)/r=\kappa$, we have $\Phi(\zeta)={\zeta}/{\kappa}+O(1)$. For strict proof, see the Supplementary Material. Thus, the geometric means enter as the kinetic factor controlling the linear growth of the Nevanlinna map at infinity. The positivity of its Nevanlinna measure then gives $\operatorname{Im}\Phi(\zeta) \geq {\operatorname{Im}\zeta}/{\kappa}$. Evaluating this inequality at $z$ gives Eq.~\eqref{eq:frequency-hierarchy}, which proves Theorem~1.

The two affinity-winding bounds \eqref{eq:affinity_winding_bounds} follow analogously from the boundary behavior of the same map near the real endpoints $\zeta=1$ and $\zeta=-1$, respectively. The corresponding limiting arguments and the rigorous derivation are given in the Supplemental Material. Thus, different analytic features of the same Floquet-Nevanlinna framework encode two complementary levels of spectral information. Its behavior at infinity retains coarse kinetic information, whereas its real-endpoint behavior recovers the universal thermodynamic localization.

\paragraph{Discussion and Conclusion ---.}
In this work, we have shown that kinetic heterogeneity imposes an additional constraint on oscillatory relaxation in a Markov cycle. The geometric means of the forward and backward rates define the factor $\kappa=(G_++G_-)/r \leq 1$, which reduces the maximal winding-resolved oscillation frequency and contracts the thermodynamically allowed spectral region. For every nonuniform cycle, $\kappa<1$, implying that the nonreal boundary of the Uhl-Seifert ellipse can be saturated only by the uniform cycle. Our result establishes an intermediate level of spectral localization between thermodynamics and complete microscopic dynamics.

The analytical framework introduced in our proof suggests a route beyond the present bound. It encodes both kinetic and thermodynamic constraints in the limiting behaviors. The large-parameter behavior encodes the geometric-mean kinetic scale, while the real-endpoint behavior yields the affinity-winding constraints. Further features of this analytic function may encode a hierarchy of kinetic and thermodynamic information, leading to progressively tighter spectral bounds. Extending our to multicyclic networks is also a natural next step, although shared transitions and coupled cycle affinities may remove the simple product structure available for the unicycle. 

\begin{acknowledgments}
We thank GPT 5.6 for assistance with analytical check and language editing. The author takes full responsibility for all mathematical statements and arguments.
\end{acknowledgments}

\bibliography{references.bib}


\makeatletter
\let\auto@bib@innerbib\CombinedSavedAutoBibInner
\makeatother

\end{document}